\documentstyle[11pt,aaspp4]{article}

\slugcomment{Accepted for publication in  Astronomical Journal}

\lefthead{Nakajima et al.}
\righthead{Star count models}

\begin{document}

\title{Infrared Star Count Models and Their Application to the Subaru
Deep Field}

\author{T. Nakajima\altaffilmark{1,5},
F. Iwamuro\altaffilmark{2},
T. Maihara\altaffilmark{3},
K. Motohara\altaffilmark{2},
H. Terada\altaffilmark{2},
M. Goto\altaffilmark{2},
J. Iwai\altaffilmark{2},
H. Tanabe\altaffilmark{2},
T. Taguchi\altaffilmark{2},
R. Hata\altaffilmark{2},
K. Yanagisawa\altaffilmark{4},
M. Iye\altaffilmark{1},
N. Kashikawa\altaffilmark{1}
and
M. Tamura\altaffilmark{1}
}

\altaffiltext{1}{National Astronomical Observatory, Mitaka, Tokyo,
181-8588, JAPAN}
\altaffiltext{2}{Department of Physics, Kyoto University,
Kitashirakawa, Kyoto 606-8502, JAPAN}
\altaffiltext{3}{Department of Astronomy, Kyoto University,
Kitashirakawa, Kyoto 606-8502, JAPAN}
\altaffiltext{4}{Okayama Astrophysical Observatory,
Kamogata, Okayama, 719-0232, JAPAN}
\altaffiltext{5}{Email: tadashi.nakajima@nao.ac.jp}


\begin{abstract}
We construct infrared star count models of the Galaxy 
applicable at faint magnitudes based on
the models and luminosity functions obtained at $V$.
There are two types of models derived in our paper.
First we derive infrared versions of the disk and
halo models by Gould et al. who obtained the local luminosity
functions and functional forms of the models
based on HST star counts (HST based models). Second we construct 
a double exponential disk model based on the luminosity function
of the nearby stars by Reid and Gizis and a halo model
based on the nearby subdwarf luminosity function of Dahn et al.
(Nearby star LF based models). In addition to
dwarfs and subdwarfs included in original models, we also
take into account L dwarfs, T dwarfs, and white dwarfs of
both disk and halo.
As a test case of the models, 
we analyze the infrared imaging data
at $J$ and $K^\prime$
obtained
during the Subaru Deep Field (SDF) survey to study stellar objects.
Out of some 350 objects, 
14 stellar objects are selected in a 
$2^\prime \times 2^\prime$ field
based on a morphological
criterion applied to the $J$ band image.
Both completeness and contamination associated with
the selection criterion are evaluated by simulations.
The $J$ band image is 57\% complete 
at $J$ = 24 and the number of contaminants is estimated
to be negligible. 
The  
prediction of 
the HST based models agrees with
the observed count at $-0.8\sigma$ and 
that of the nearby star LF based models
also agrees with
the observations at $+1.0\sigma$.
The observed count is between the predictions of the two types of models that 
have contradictory local luminosity functions. 
With our limited statistics, the observing data do not
favor a particular type of models.
Infrared star count models we have obtained predict that
the Next Generation Space Telescope will primarily see
T dwarfs, M subdwarfs, and old halo white dwarfs at faint magnitudes.

\end{abstract}




%

\section{Introduction}

Deep infrared images have been obtained so far for the purpose of 
deep galaxy counts (e.g. \cite{bers98}). It is indeed the case that
the majority of the objects detected in those images are galaxies
and stellar objects are the minority.  
Infrared images of 
the Subaru Deep Field (SDF), which we are going to analyze in this paper,
were obtained also for the purpose of a galaxy count analysis 
(\cite{mai00}). However as we show later,
these images contain interesting information about faint
stellar/substellar populations, since they reach the magnitude limits
which no stellar astronomer has ever experienced.

The SDF images were obtained at $J$ and $K^\prime$.
Once stellar objects are selected from a catalog of detected sources
by a criterion on the morphology of individual sources,
only remaining information is color-magnitude information.
As an aid to interpret the color-magnitude information, 
we extensively utilize Galaxy model predictions on the stellar/substellar
populations. 
Infrared star count models at faint magnitudes have not been
constructed previously. We generate models for $J$ 
based on those constructed for
$V$ and color-magnitude relations for visual and infrared.
We also take into account faint stellar/substellar 
populations such as L dwarfs, T dwarfs, and white dwarfs
of the Galactic disk and halo.
We compare the model predictions with observations
and examine the consistency. 
We also utilize these models to predict the Galaxy 
observed by the Next Generation Space Telescope (NGST).

In the past, infrared images as deep as $K$ = 19 were
analyzed from the point
of view of stellar astronomy. 
\cite{hu94} studied
the deep imaging data obtained for a deep galaxy count and
concluded that very little of dark matter in the Galactic halo
can be made of low-mass hydrogen-burning stars. 
In our small field of view,
there are only 14 stellar objects.
However analyses of a small number of stellar objects have been
already done in the Hubble Deep Field (\cite{els96}, \cite{men96}). 
Using 
different criteria,
Elson et al. found 59 unresolved objects while Mendez et al. selected
only 14 stellar objects.  Elson et al. counted only nine objects as
candidates for low mass halo stars and ruled out a steeply
rising luminosity function fainter than $M_V$ = 12.
Mendez et al. found that the luminosity function
for halo objects must be a factor of 2 smaller than
that predicted by an extrapolation of the solar neighborhood
luminosity function for disk stars.

Star count models are constructed in \S2.
The reduction and analysis 
of the SDF data are described in \S3.
Model predictions are compared
with observations in \S4. Individual objects in SDF
are discussed in \S5 and
the Galaxy seen by NGST is
predicted in \S6. 

\section{Star Count Models}

Infrared star count models applicable to very faint magnitudes
do not exist in the literature. 
So we construct them, starting
from the models obtained in the visible, using  color-magnitude
relations. 
We also use a $K$ band luminosity function for the disk as a starting
point when available. We obtain color-magnitude, magnitude-magnitude
relations empirically from published photometric data when necessary.
We first construct models by Gould et al. who used HST star count
data. Their local luminosity functions (LLFs)
of the disk and
halo obtained by fitting distant stars are inconsistent
with the LLFs of nearby stars. We also construct
models different from those of Gould et al.,
using the nearby star LLFs.

\subsection{Model A: Disk model by Gould et al.}

We first construct an infrared version of the disk 
model by (\cite{gou97}) who fitted the disk structure parameters
and the LLF to star count data from
HST observations of high-latitude fields.
The distribution of stars is modeled as a function of Galactic
position and absolute magnitude by

\begin{equation}
\Phi(M_V,z,R) = \Phi(M_V) \nu(z) \exp\bigl(-\frac{R-R_0}{H}\bigr),
\end{equation}

where $\Phi(M_V)$ is the LLF, $(R,z)$
is the Galactic position in cylindrical coordinates, 
$R_0$ = 8 kpc is the Galactocentric distance, and $H$ = 2.92 kpc is
the disk scale length. The vertical distribution function
is given by 

\begin{equation}
\nu(z) = (1-\beta){\rm sech}^2\frac{z}{h_1} + \beta\exp\frac{-|z|}{h_2},
\end{equation}

where $h_1$ = 320 pc, $h_2$ = 643 pc, and $\beta$ = 21.6\%.

The local disk/thick disk ratio, $\beta$, adopted by Gould et al.
is more than a factor of two higher than any other estimates
and this discrepancy is reflected in the very small scale height,
$h_2$. This small scale height results in a smaller star count
estimate at faint magnitudes.
The LF is given  for main-sequence
stars with $8.0 < M_V  < 18$. Since we are interested in
star counts at faint magnitudes, this range of the LF is
appropriate.  
The infrared color-magnitude relation is not
available for thick disk stars
and we use the color-magnitude information of thin disk stars 
in \cite{leg92} for the entire disk stars to calculate
star counts in the infrared. 
\cite{rei97} have found that 
the metallicity of the stars do not change significantly
out to the scale height of 2 kpc. This may be a support
for the use of the thin-disk color-magnitude relation
for the entire disk stars.

\subsection{Model B: Halo model by Gould et al.}

We also construct an infrared version of the halo model by
\cite{gou98} who fitted the halo structure parameters
and the LLF to 
HST star count data.
The distribution of halo stars is given by

\begin{equation}
\Phi(M_V,R,z) = \Phi_h(M_V)\Bigl[\frac{R^2+(z/c)^2}{R_0^2}\Bigr]^{-l/2},
\end{equation}

where $\Phi_h(M_V)$ is the LLF, $c$ = 0.82 is the flattening
parameter, and $l$ = 3.13 is the power.
There is no infrared color-magnitude relation available
for halo stars.
To calculate star counts at $I$, $J$, $K$ and $J-K$ color, following
empirical relations were obtained by fitting
extreme M subdwarf data in \cite{lah98};

\begin{equation}
M_I = 0.745 M_V + 1.122,
\end{equation}

\begin{equation}
M_J = 0.638 M_V + 1.249,
\end{equation}

\begin{equation}
M_K = 0.649 M_V + 0.503,
\end{equation}

\begin{equation}
J-K = -0.011 M_V + 0.747.
\end{equation}

The sample in Leggett et al. also includes a small number of
M subdwarfs (not extreme M subdwarfs), but they are not used in the fitting.

\subsection
{Model C: Exponential disk model based on nearby star LF}



We construct a double exponential disk model using
the nearby star $K$ band LLF by \cite{rg97}.
The distribution of disk stars is given by

\begin{equation}
\Phi(M_K,R,z) = \Phi(M_K)\exp(\frac{R_0-R}{H})
                \Bigl[(1-\beta)\exp(-\frac{|z|}{h_1})
                        +\beta\exp(-\frac{|z|}{h_2})\Bigr],
\end{equation}

where $\Phi(M_K)$ is the $K$ band LLF, 
$R_0$ = 8 kpc is the Galactocentric distance,
$H$ = 3.5 kpc is the
Galactic scale length, $h_1$ = 325 pc is the scale height
of thin disk stars. For the scale height $h_2$ and
the normalization $\beta$ of the thick disk, we consider
the combinations of  ($h_2$,$\beta$) = (910 pc, 6\%) 
(Model C1: \cite{bus99}) and ($h_2$,$\beta$) =
(1500 pc,2\%) (Model C2: \cite{rei93}). 
Using the color-magnitude relation of thin disk stars by
\cite{leg92}, the model is constructed so that it predicts
star counts at $V$, $I$, $J$ and $K$.
We note that our model has a limitation in the use of
the thin disk color-magnitude relation for the thick disk.
However the effect of metallicity on the $J-K$ color is relatively
small and the $J$ band count is not significantly affected by
this problem.

\subsection{Model D: Halo model based on nearby star LF} 

The model D is based on the LLF of nearby stars by \cite{dah95}.
The density distribution of halo stars is given by

\begin{eqnarray}
\Phi(M_V,R,z) & = & \Phi(M_V)\Bigl(\frac{x}{R_0}\Bigr)^{-7/8}
\Bigl\{\exp[-10.093(\frac{x}{R_0})^{1/4}+10.093]\Bigr\} \nonumber \\
& & \times \Bigl[1-0.08669/(\frac{x}{R_0})^{1/4}\Bigr],
\end{eqnarray}

where $x = (R^2 + (z/c)^2)^{1/2}$ and c = 0.85 (\cite{rei93}).
$\Phi(M_V)$ of Dahn et al. is significantly greater than
the LLF of \cite{gou98}.
Star counts at $I$, $J$, and $K$ are estimated from the magnitude-magnitude
relations eqs (6), (7), and (8).

\subsection{L dwarfs, T dwarfs and white dwarfs}

Dwarfs later than M, namely L and T dwarfs (\cite{kir99})
are also included in our models. The local number density of L dwarfs
is estimated by the 2MASS observations of Kirkpatrick et al. 
to be 7.2$\times 10^{-3}$ pc$^{-3}$ using 1/V$_{\rm max}$ estimates. 
The local density
of T dwarfs is estimated by the SDSS observations of \cite{str99} and
\cite{tsv00} to be 1.4$\times 10^{-1}$ pc$^{-3}$. This
assumes
that the survey volume extend out to 10 pc.
As for T dwarfs, we note that
the 2MASS observations
by \cite{bur99} give a lower number density than
the SDSS observations.
For simplicity, L dwarfs are assumed to be a single population
with $M_{J}$ = 13.0 and $J-K$ = 1.5, and T dwarfs 
with $M_{J}$ = 15.4 and $J-K$ = -0.1 like Gl 229B
(\cite{mat96}). The ages of L and T dwarfs are expected to
be order of 1 Gyr and they belong to the young disk population.
The scale heights of both L and T dwarfs
are assumed to be 200 pc or the scale height of F stars.

We then consider white dwarfs. For hot white dwarfs, we compute
the LF by \cite{lie88}
at V. Since there is no good observational color-magnitude relation 
for hot white dwarfs usable to derive an infrared luminosity function,
we are satisfied with the V band model to order estimate infrared
star counts.
The luminosity function of cool white dwarfs ($M_{bol} > 13$)
has been obtained by \cite{leg98} for $M_{bol}$ based on
visual and infrared observations. By fitting her observational
data, we have obtained following relations among  $M_{bol}$, 
$M_{V}$, $M_{I}$, $M_{J}$, $M_{K}$ and $J-K$;

\begin{equation}
M_V = 1.103 M_{bol} - 1.311,
\end{equation}

\begin{equation}
M_I = 0.743 M_{bol} + 3.084,
\end{equation}

\begin{equation}
M_J = 0.542 M_{bol} + 5.593,
\end{equation}

\begin{equation}
M_K = 0.494 M_{bol} + 6.041,
\end{equation}

\begin{equation}
J-K = 0.047 M_{bol} - 0.448.
\end{equation}

Using these relations, we include cool white dwarfs in the disk models.
It turned out that disk white dwarfs do not have
significant contribution at faint magnitudes.

We also consider old white dwarfs in the halo (\cite{hod00}).
These white dwarfs are very faint and blue in the infrared.
The absolute magnitudes are assumed to be of WD 0346+246 
(\cite{hod00}) and the number density is assumed to be
7$\times 10^{-4}$ pc$^{-3}$ (\cite{iba00}). 
These old halo white dwarfs turn out to be important in
the star count by NGST at the faintest magnitudes as will
be discussed in \S6.

\section{Reduction and Analysis of the SDF data}

Details of
the observations of the Subaru Deep Field (SDF) are described in
\cite{mai00}.  The SDF is a 2$^\prime \times 2^\prime$
field
centered on
($\alpha$,$\delta$) = ($13^h24^m21.3^s$,
27$^\circ29^\prime23^{\prime\prime}$) (J2000)
or (l,b) = ($35.\hspace{-2pt}^\circ7,82.\hspace{-2pt}^\circ2$).  
For the purpose of star-galaxy separation, exposure frames
taken only under good seeing conditions were coadded to
form final images. The total integration time of
the coadded frames at $J$ is 
4.97 h out of 12 h of observations   
and that at $K^\prime$ is  4.13 h out of 10 h of observations. 
The FWHMs
of the resultant PSFs are $0.\hspace{-2pt}''39$ and $0.\hspace{-2pt}''29$
respectively at $J$ and $K^\prime$.


A source catalog  was created for each band
by the SExtractor (\cite{bert96}).
For each source, its position, magnitude, FWHM of the image,
and `stellarity' were recorded. 
The stellarity is a parameter defined in the SExtractor
which quantifies how stellar the object morphology is.
It is unity
for a completely stellar object and zero for a completely
non-stellar object. The stellarity is an output parameter of
a neural network. The purpose of
the use of the neural network is to provide an optimal transformation
from the parameter space defined by a set of observables describing
the object, to the one dimensional space of stellarity.
Input parameters are eight isophotal areas, the peak intensity, and
the seeing FWHM. 

Star-galaxy separation is a critical process in the analysis
of a deep image. 
Its difficulty in a deep optical star count has been mentioned by
\cite{rei96}, who analyzed visible images obtained 
on the Keck telescope under $0.\hspace{-2pt}''5$ seeing. They conclude that
the decreasing average angular size of galaxies with fainter
magnitude limits star-galaxy separation to $R$ $<$ 25.5 mag.
In separating stellar objects from galaxies, we
impose a rather strict quantitative criterion for the
selection of stellar objects and separate stars and galaxies.
To evaluate the objectiveness of the selection criterion,
we simulate images with different sizes at different
magnitudes and apply the selection criterion to estimate
the fraction of stars classified as galaxies (incompleteness)
and that of galaxies classified as stars (contamination).
So the incompleteness and contamination are obtained
as functions of image size and magnitude. 
Then we apply an incompleteness factor for each
stellar object to estimate the true star count and
apply a contamination factor for each galaxy
to evaluate the number of contaminating galaxies.

\subsection{Selection Criterion}

We first simulate 500 PSFs with FWHM of $0.\hspace{-2pt}''39$ per 0.25 mag for
$J$ and 500 PSFs with FWHM of $0.\hspace{-2pt}''29$ for 
$K^\prime$ between 16 and 26 mag.
Simulated objects are meant to be stars.
Then we run the SExtractor to detect each artificial
source and measure its magnitude, FWHM and stellarity. 

The details of the simulation procedure for each SExtractor run 
are as follows. 
We randomly place on the true object image 50 
artificial objects with random counts
within  the magnitude range between m and m+0.25.
We run the SExtractor using the same detection conditions as for
the true objects and compare the resultant catalog with the
true object catalog to create a list of new objects which newly appeared
or brightened by more than 0.75 mag. We compare the coordinates
of the 50 artificial objects placed prior to the SExtractor run with those 
of the resultant new objects and pick up the nearest neighbor pairs 
within 3 pixels ($0.\hspace{-2pt}''35$) as recovered artificial objects.
We repeat the same procedure ten times and move to the next magnitude bin.
The number of the artificial objects, 50 is smaller than the total
number of true objects, 350, and the confusion in the
recovery process  is due to the true objects.
If 500 artificial objects were placed at once, confusion among 
them would dominate. We have avoided this situation by
limiting the number to 50 for each SExtractor run, which is
still much more efficient than the ideal case of one at a time.

At $J$, the FWHMs of
detected objects stay near $0.\hspace{-2pt}''39$ from 16 to 22 mag,
grow gradually from 22 to 23.5 and start to diverge at 24 mag.
Below 25.5 mag, most of the sources are undetected.
At $K^\prime$, the FWHMs of detected objects stay close to
$0.\hspace{-2pt}''29$ from 16 to 21 mag, grow slowly from 21 to 22.5 mag,
and diverge at 23 mag. Below 24.5 mag, most of the sources
are undetected. 
At $J$, the stellarity stays close to unity from 16 to 23 mag,
and then gradually drops below 0.8 at fainter magnitudes. 
Below 24.5 mag, only a few sources
have the stellarity above 0.8.
At $K^\prime$, the stellarity stays close to unity from 16 to 22 mag,
and gradually drops below 0.8 at fainter magnitudes.
Below 24 mag, only one object has the stellarity above 0.8.  
From these simulations, 
we have found that $J$ is more sensitive than 
$K^\prime$  for the sources with $J-K^\prime$ $<$1.0.
In selecting stellar objects, we use the $J$ band image, and
the $K^\prime$ band image is used to measure the color. 

We further simulate 500 PSFs per 0.25 mag with FWHMs of 1.26$\times$seeing,
$1.26^2 \times$seeing, and $1.26^3 \times$seeing between 16 and 26 mag
for $J$. These sources are meant to simulate non-stellar objects.
We again run SExtractor to detect each artificial source and measure
its magnitude, FWHM and stellarity.
Simulated stellar objects are well separated from simulated
non-stellar objects in FWHM down to 24 mag.  
Below $J$=24,
stellar and
non-stellar
objects tend to mix toward fainter magnitudes.
The stellarity stays above 0.8 at magnitudes brighter than
$J$=17, 18, and 21
respectively 
for PSFs with FWHMs of $1.26^3 \times$, $1.26^2 \times$
and  1.26$\times$seeing, and goes down near zero and comes
back up to 0.7 below 24 mag.  
At brighter magnitudes, the FWHM of each artificial source is a good
discriminator between a stellar and non-stellar object.
At fainter magnitudes, the stellarity appears to be a superior
discriminator
because that of a non-stellar object rarely becomes above 0.7.

From the simulations so far, we find following conditions for FWHM and
stellarity to separate stellar objects from non-stellar objects
for $J$.
Stellar objects satisfy

\begin{equation}
{\rm FWHM}(^{\prime\prime}) \leq 0.47 + 10^{0.4(J -26)},
\end{equation}

and 

\begin{equation}
{\rm Stellarity} \geq 0.8.
\end{equation}

The performance of the selection conditions is graphically shown
in Figure 1.

\placefigure{fig1}

\subsection
{Selected Stellar Objects}

From the selection criterion given in
the previous section, 14  objects are
selected (Table 1).
$J$ and $K^\prime$ images of each object is shown in Figure 2.
The completeness correction and the number of contaminants
are estimated and given in Table 2. 
From Table 2, the star count is more than 94\% complete
above $J$ = 22.5 and 57\% complete above $J$ = 24.
The completeness is only 28\% between $J$ = 24 and 24.5.
After the completeness correction, the estimated total
count is 24.3. 
The total number of contaminants is only 0.03 and is negligible.
This indicates how strict the selection criterion is. 
The color-magnitude diagram of the stellar objects is given
in Figure 3. Above $J$ = 24, the $J-K^\prime$ colors of eleven objects
are between 0.40 and 0.72. However below $J$ = 24, one object is
very blue (0.0) and two others are very red ($>$ 1.0).

\placefigure{fig2}

\placetable{table1}

\placefigure{fig3}

\placetable{table2}

\section{Comparison of observations and models}

The completeness of the observed star count is 57\% at $J$ = 24.
Model predictions and the completeness corrected counts
are compared in Tables 3 and 4.
The $J-K^\prime$ 
color of observed eleven objects brighter than $J$ = 24 is confined
in the range from 0.4 to 0.7.  If the completeness correction
is applied, the effective star count brighter than $J$ = 24 is 13.5.

The disk model A predicts only 2.6 stars brighter than
$J$ = 24, 
whose $J-K$ color  ranges from 0.78 to 0.96. 
We note that
the color-magnitude relation is of the thin disk. 

The halo model B predicts 7.8 halo stars.  Combined
with the disk model A, the models
by Gould et al. based on HST observations
predict 10.4 stars in the magnitude range
between J = 18 and 24, which agrees
with  the completeness corrected count at $-0.84\sigma$.

\placetable{table3}

The disk model C1 (thick disk scale height 910 pc) predicts
5.5 stars with $J-K$ between 0.87 and 0.96, while
the disk model C2 (thick disk scale height 1.5 kpc) predicts
8.2 stars with $J-K$ between 0.85 and 0.96.
For both models, the thin disk contribution is small, 1.9 stars.

The halo model D predicts 11.8 halo stars within the color range
of 0.59 and 0.65.  If some halo stars have colors of
M subdwarfs instead of those of extreme M subdwarfs, the observed
stars brighter than $J$ = 24 may all be halo stars.
In combining the disk model with the halo model D,
the model C1 is preferred because of the smaller predicted number
count. 
The total predicted star count of C1$+$D 
is 17.4 which is slightly greater than the completeness corrected count of 13.5
by 1.0 $\sigma$.
The observed star count is between the prediction of
models based on HST data by
Gould et al. and that of models based on nearby stars LFs.
Our limited statistics do not have strong constraint 
on the discrepancy between the results of two types of models.

\placetable{table4}

Apart from stars, the disk models predict a small number of 
substellar objects brighter than $J$ = 24. 
The disk model A and C estimate the number
of T dwarfs to be 1.4 and 0.39 respectively. Combining the local number density
of T dwarfs with the Galactic structure
of disk model A of Gould et al. apparently results in
overestimating the number of T dwarfs.
The disk model C is preferred for the combination with the local
number density. 
The disk model A and C predict the number of L dwarfs to be
0.65 and 0.1. Since no object as red as L dwarfs 
brighter than $J$=24 was detected,
the Galactic structure of the disk model C is again preferred.

\section{Individual Objects}

\subsection{Blue M subdwarfs, \#7 and LHS 1826}

The object \#7 is 
bluer ($J-K^\prime$ = 0.4) than any other object with $J<24$. 
It is also bluer than any of the nearby extreme subdwarfs whose colors
are given in \cite{lah98}.  According to the calculations by
\cite{sau94} of atmospheres of zero-metal low-mass objects,
the $J-K$ color of a zero-metal object near the stellar/substellar
boundary will be blue and 
negative due to H$_2$ collision induced absorption (CIA).
The color of \#7 may be affected by H$_2$ CIA significantly.
H$_2$ CIA most likely affects the coolest
and most-metal-poor object.  In the solar neighborhood, the coolest
and most-metal-poor object is LHS 1826 (\cite{giz97}) whose infrared
color was previously unknown. We obtained $JHK$ photometry of LHS 1826
at the Okayama Astrophysical Observatory on 1999, November 30,
using the facility infrared
imager, OASIS (\cite{oku00}). 
Resultant magnitudes are $J$ = 15.70 $\pm$0.04, 
$H$ = 15.43 $\pm$
0.05, and $K$ = 15.32 $\pm$ 0.07. Therefore $J-K$ = 0.36 $\pm$ 0.08 and
it is as blue as \#7.  It is important to measure
the parallax of LHS 1826 and obtain its absolute magnitude.
It may be the least luminous extreme M subdwarf known so far.
\#7 and LHS 1826 indicate the presence of
a faint blue halo population previously unnoticed. 

\subsection{Red objects, \#11 and \#12}

Errors in $J-K^\prime$ colors of the objects with $J > 24$
are large. However, it is very likely that
at least either \#11 or \#12 has $J-K^\prime > 1.0$.
Only L dwarfs are the stellar/substellar objects as red as these, 
but the expected
number of L dwarfs between J = 24 and 25 predicted by the
model C1 is 0.06. 

Now we consider the possibility that at least one of the objects is 
extragalactic, namely an AGN. 
In the local universe, the fraction of AGNs to the entire galaxy
population is estimated to be 1.3\% based on the data of the CfA redshift
survey
(\cite{huc92}). There are some 350 galaxies in SDF. If the
fraction of AGNs does not evolve, there may be four AGNs in the field.
However, we know that galaxies evolve and so do quasars (e.g. \cite{sch95}). 
So the validity of the number estimate based on the local universe
information is questionable. 

PG quasars are red and most of them have $J-K$ between 1 and 
2 (\cite{neu87}). In terms of colors, low-redshift quasars are
candidates. 
However they are extremely bright ($J<$12).
If we assume that the object has a power-law continuum
whose spectral index $\alpha = -0.653$ is consistent with
$J-K^\prime$ = 1.5, the combinations of absolute $B$ magnitude,
$M_B$ and z can be obtained for the apparent magnitude of $J$ = 24.2.
Here the cosmological parameters we assume are $H_0$ = 70 kms$^{-1}$
Mpc$^{-1}$, $\Omega_m$ = 0.3, and  $\Omega_\lambda$ = 0.7.
Allowed combinations of ($M_B$,z) are (-22,1.2), (-24,2.9), and
(-26,7.0). So if the object is a quasar with $M_B$ = -26,
it must be located at very high redshift. If such quasars
are as numerous as to be detected in a small field,
the number density may be high enough to 
affect reionization of the universe.

\section{The Galaxy Seen by NGST}

Since we have obtained the star count models which are capable
of predicting the appearance of the Galaxy at faint magnitudes,
it is interesting to use them to envision the Galaxy seen
by NGST. In Table 5, surface number densities (arcmin$^{-2}$) of objects
between $K$ = 27 and 30 are given for the north Galactic pole.

In the galactic disk, T dwarfs are by far the most dominant
population.
However the expected number count is small ($<0.01$) due to
the small vertical scale height.
There may be fainter and bluer M subdwarfs than predicted,
because the LFs in the models extend only to $M_V$ = 14, while
those of the disk extend to $M_V$ = 18. Old halo white dwarfs
are the most numerous of all stellar populations.
T dwarfs are blue in the near infrared due to molecular absorption
bands. Cool M subdwarfs and halo white dwarfs are blue due to 
H$_2$ CIA. Stellar astronomy will have a blue faint end in the near infrared.

\placetable{table5}

\section{Conclusions}

Our primary conclusions follow.

(1)
We constructed infrared star count models of the Galaxy 
applicable at faint magnitudes based on
the models and luminosity functions obtained at $V$.
There are two types of models derived in our paper.
First we  derived infrared versions of the disk and
halo models by Gould et al. who obtained the local luminosity
functions and functional forms of the models
based on HST star counts. 
Second we constructed 
a double exponential disk model based on the luminosity function
of the nearby stars by Reid and Gizis and a halo model
based on the nearby subdwarf luminosity function of Dahn et al.. 
In addition to
dwarfs and subdwarfs included in original models, we also
took into account L dwarfs, T dwarfs, and white dwarfs of
both disk and halo.

(2)
As a test case of the models, 
we  analyzed the infrared imaging data
at $J$ and $K^\prime$
obtained
during the Subaru Deep Field (SDF) survey to study stellar objects.
Out of some 350 objects, 
14 stellar objects were selected in a 
$2^\prime \times 2^\prime$ field
based on a morphological
criterion applied to the $J$ band image.
Both completeness and contamination associated with
the selection criterion were evaluated by simulations.
The $J$ band image is 57\% complete 
at $J$ = 24 and the number of contaminants was estimated
to be negligible. 
The  
prediction of 
the HST based models agrees with
the observed count at $-0.8\sigma$ and 
that of the nearby star LF based models
also agrees with
the observations at $+1.0\sigma$.
The observed count is between the predictions of the two types of models that 
have contradictory local luminosity functions. 
With our limited statistics, the observing data do not
favor a particular type of models.

(3)
Infrared star count models we obtained predict that
the Next Generation Space Telescope will primarily see
T dwarfs, M subdwarfs, and old halo white dwarfs at faint magnitudes.

\acknowledgments

We thank the staff of the Subaru Observatory for making
the SDF observations possible. We thank the anonymous referee
for useful comments on the manuscript.
TN is supported by Grant-in-Aid for Scientific Research of
the Japanese Ministry of Education, Culture, Sports, and
Science (No. 10640239).

\clearpage

\begin{deluxetable}{cccccccc}
\footnotesize
\tablecaption{Stellar Objects \label{table1}}
\tablewidth{0pt}
\tablehead{
\colhead{\#} & \colhead{($\alpha,\delta$)(J2000)}   
& \colhead{$J$}   & \colhead{$\sigma(J)$} & 
\colhead{FWHM ($J$)}  & \colhead{Stellarity} & \colhead{$J-K^\prime$} & 
\colhead{$\sigma(J-K^\prime$)} 
} 
\startdata
 1& 132229.6+272841.3  & 21.999 & 0.016 & 0.420 & 0.990 & 0.518 &0.04 \nl 
2& 132240.0+272932.1  &18.217& 0.001 & 0.435 &0.980 & 0.696 &0.0014\nl
3& 132250.6+272900.9  & 20.209 & 0.003 & 0.418 & 1.000 & 0.714 &0.006 \nl
4& 132304.3+272904.3  & 20.269 & 0.004 & 0.416 & 0.980 & 0.523 &0.008  \nl
5& 132309.5+272949.8  &21.389& 0.009 & 0.435 &0.980 & 0.678 &0.014  \nl
6& 132314.1+272941.9  &20.869& 0.006 & 0.421 &0.980 & 0.635 &0.0116 \nl
7& 132315.8+272921.2  & 23.545 & 0.048 & 0.416 & 0.960 & 0.403 &0.10  \nl
8& 132344.9+272841.3  &22.191& 0.021 & 0.461 &0.970 & 0.669 &0.039   \nl
9& 132357.7+272951.8  &21.963& 0.014 & 0.419 &0.980 & 0.473 &0.030  \nl
10& 132525.4+272922.4 & 23.886 & 0.057 & 0.386 & 0.970 & 0.661 &0.115 \nl
11& 132558.9+272854.2 & 24.215 & 0.121 & 0.423 & 0.890 & 1.288 &0.17  \nl
12& 132607.4+272907.0 & 24.433 & 0.115 & 0.474 & 0.870 & 1.648 &0.16  \nl 
13& 132617.2+272945.4 &23.804& 0.066 & 0.501 &0.950 & 0.721 &0.141  \nl
14& 132622.1+273010.3 &24.176& 0.109 & 0.458 &0.810 & 0.002 &0.255 \nl

\enddata
 
\end{deluxetable}

\clearpage

\begin{deluxetable}{cccc}
\footnotesize
\tablecaption{Completeness correction and contamination \label{table2}}
\tablewidth{0pt}
\tablehead{
\colhead{$J$} & \colhead{Detection}   
& \colhead{Corrected Detection}   & \colhead{Contamination} 
} 
\startdata
   17.2500 &    0.0000 &    0.0000 &    0.0000\\
   17.7500 &    0.0000 &    0.0000 &    0.0000\\
   18.2500 &    1.0000 &    1.0142 &    0.0000\\
   18.7500 &    0.0000 &    0.0000 &    0.0000\\
   19.2500 &    0.0000 &    0.0000 &    0.0000\\
   19.7500 &    0.0000 &    0.0000 &    0.0000\\
   20.2500 &    2.0000 &    2.0555 &    0.0000\\
   20.7500 &    1.0000 &    1.0277 &    0.0000\\
   21.2500 &    1.0000 &    1.0320 &    0.0000\\
   21.7500 &    2.0000 &    2.0725 &    0.0000\\
   22.2500 &    1.0000 &    1.0638 &    0.0000\\
   22.7500 &    0.0000 &    0.0000 &    0.0000\\
   23.2500 &    0.0000 &    0.0000 &    0.0020\\
   23.7500 &    3.0000 &    5.2724 &    0.0180\\
   24.2500 &    3.0000 &   10.7527 &    0.0100\\
   24.7500 &    0.0000 &    0.0000 &    0.0000\\

\enddata
\end{deluxetable}

\clearpage

\begin{table*}
\begin{center}
\begin{tabular}{ccccc}
\tableline\tableline
$J$  &  disk model A  & halo model B & A+B & Corrected count   \\
\hline
18-20 &  1.53  & 0.28  & 1.82  &   1.01$\pm$1.00\\
20-22 &  0.85  & 1.86  & 2.72  &   6.19$\pm$2.49 \\   
22-24 &  0.23  & 5.68  & 5.91  &   6.34$\pm$2.51  \\
\tableline
Total & 2.61  &  7.82  &  10.45($-0.84\sigma$) &   13.54$\pm$3.68 \\
\tableline
\end{tabular}

\end{center}

\caption{Models by Gould et al. \label{table3}}

\end{table*}

\clearpage

\begin{table*}
\begin{center}

\begin{tabular}{ccccc}
\tableline\tableline
J  &  disk   model C1  & halo model D & C1+D & Corrected count   \\
\hline
18-20 &  2.13  & 0.40  & 2.74  &   1.01$\pm$1.00\\
20-22 &  2.32  & 3.00  & 5.31  &   6.19$\pm$2.49 \\   
22-24 &  1.08 &  8.43  & 9.50  &   6.34$\pm$2.51  \\
\tableline
Total & 5.53  &  11.83  &  17.36($1.04\sigma$) &   13.54$\pm$3.68 \\
\tableline
\end{tabular}

\end{center}

\caption{Models based on nearby star LFs \label{table4}}

\end{table*}

\clearpage

\begin{table*}
\begin{center}

\begin{tabular}{ccccc}
\tableline\tableline
Stellar population & disk model A & halo model B  & disk model C  &
halo model D    \\
\hline
T dwarfs          & 0.002 & $---$   &  0.005  &  $---$    \\
M subdwarfs       &  $---$ & 0.50  &    $---$  &  0.12    \\   
Halo white dwarfs & $---$  & 1.53  &    $---$  &  1.23    \\
\tableline
\end{tabular}

\end{center}

\caption{Predicted surface number density (arcmin$^{-2}$) \label{table5}}

\end{table*}

\clearpage

%
%

\clearpage

\figcaption[figure1.ps]{Selection of stellar objects.
In the upper panel, small dots indicate simulated objects
for FWHMs $0.\hspace{-2pt}''39$, 
$1.26\times0.\hspace{-2pt}''39$, 
$(1.26)^2\times0.\hspace{-2pt}''39$, 
and $(1.26)^3\times0.\hspace{-2pt}''39$.
500 objects are simulated for each 0.25 mag interval.
Dots generated for different FWHMs tend to mix at fainter 
magnitudes. Observed objects with different J magnitudes
are indicated by circles with various sizes.
The selection criterion for FWHM is shown by the solid line
below which the sources are regarded as stellar objects.
In the lower panel, the stellarity condition is imposed on
both observed and simulated objects
and only those that satisfy the condition are plotted.
When both the FWHM and stellarity conditions are imposed,
the mixing of dots generated for different FWHMs
is fully suppressed.  \label{fig1}}

\figcaption[figure2.ps]{$J$ and $K^\prime$ images of stellar objects (The $J$ image 
 is on top of 
the $K^\prime$ image in each pair).
Pairs of \#1 through \#5 are presented in the bottom raw
from left to right, those of \#6 through \#10 in the mid raw
and those of \#11 through \#14 in the top raw. \label{fig2}}

\figcaption[figure3.ps]{$J$ vs $J-K^\prime$ diagram. 
All the eleven objects with $J<$24 have
$J-K^\prime$ colors in the range of 0.4 and 0.7. All the objects
with $J>$ 24 have exotic colors.  
Error bars are of 1 $\sigma$.
\label{fig3}}


\begin{thebibliography}{}
\bibitem[Bershady et al. (1998)]{bers98} Bershady, M. A.,
Lowenthal, J. D. and Koo, D. C. 1998, \apjl, 522, L15
\bibitem[Bertin and Arnouts (1996)]{bert96} Bertin, E. and 
Arnouts, S. 1996, \aap, 117, 393
\bibitem[Burgasser et al. (1999)]{bur99} Burgasser, A. J.,
Kirkpatrick, J. D., Brown, M. E., Reid, I. N.,
Gizis, J. E., Dahn, C. C., Monet, D. G., Beichman, C. A., Liebert, J.,
Cutri, R. M., and Skrutskie, M. F. 1999, \apjl, 522, L65
\bibitem[Buser et al. (1999)]{bus99}
Buser, R., Rong, J., and Karaali, S. 1999, \aap, 348, 98
\bibitem[Dahn et al. (1995)]{dah95} Dahn, C. C., Liebert, J., 
Harris, H. C., and Guetter, H. H. 1995,
The Bottom of the Main Sequence And Beyond, ed. Tinney C. G.
(Berlin: Springer) p239
\bibitem[Elson et al. (1996)]{els96} Elson, R. A. W., Santiago, B. X.,
and Gilmore G. F. 1996, NewA, 1, 1
\bibitem[Gizis and Reid (1997)]{giz97} Gizis, J. E. and Reid, I. N.
1997, \pasp, 109, 849
\bibitem[Gould et al. (1997)]{gou97}
Gould, A., Bahcall, J. N., and Flynn, C. 1997, ApJ, 482, 913
\bibitem[Gould et al. (1998)]{gou98}
Gould, A., Flynn, C., and Bahcall, J. N. 1998, ApJ, 503, 798
\bibitem[Hodgkin et al. (2000)]{hod00}
Hodgkin, S., Oppenheimer, B., Hambly, N., Jameson, R., Smartt, S.,
and Steele, I. 2000, Nature, 403, 57
\bibitem[Hu et al. (1994)]{hu94}
Hu, E. M., Huang, J. S., Gilmore, G., and Cowie, L. L. 1994, Nature,
371, 493
\bibitem[Huchra and Burg (1992)]{huc92}
Huchra, J. P., and Burg, R.  1992, \apj, 393, 90
\bibitem[Ibata et al. (2000)]{iba00}
Ibata, R., Irwin, M., Bienayme, O., Scholz, R., and Guibert J.  2000
\apjl, 532, L41
\bibitem[Kirkpatrick et al. (1999)]{kir99}
Kirkpatrick, J. D., Reid, I. N., Liebert, J., Cutri, R. M., Nelson, B.,
Beichman, C. A., Dahn, C. C., Monet, D. G., Gizis, J. E., and
Skrutskie, M. F.
1999, \apj, 519, 802
\bibitem[Leggett (1992)]{leg92}
Leggett, S. K. 1992, ApJS, 82, 351
\bibitem[Leggett (1998)]{leg98}
Leggett S. K. 1998, ApJ, 497, 294
\bibitem[Leggett et al. (1998)]{lah98}
Leggett, S. K., Allard, F., and Hauschildt, P. H. 1998, ApJ, 509, 836
\bibitem[Liebert et al. (1988)]{lie88}
Liebert, J., Dahn, C. C., and Monet, D. G. 1988, ApJ, 332, 891 
\bibitem[Maihara et al. (2000)]{mai00}
Maihara, T. et al. 2000, submitted to PASJ
\bibitem[Matthews et al. (1996)]{mat96}
Matthews, K., Nakajima, T., Kulkarni, S. R., and Oppenheimer, B. R.  1996,
AJ, 112, 1678
\bibitem[Mendez et al. (1996)]{men96}
Mendez, R. A., Minniti, D., Marchi, G. De, Baker, A., Couch, W. J. 1996,
MNRAS, 283, 666
\bibitem[Neugebauer et al. (1987)]{neu87} Neugebauer, G., Green,
R. F., Matthews, K., Schmidt, M., Soifer, B. T., and Bennet, J. 1987,
ApJS, 63, 615
\bibitem[Okumura et al. (2000)]{oku00}
Okumura, S., Nishihara, E., Watanabe, E., Mori, A, Kataza, H, and Yamashita,
T. 2000, submitted to PASJ
\bibitem[Reid and Majewski (1993)]{rei93}
Reid, I. N.,  and Majewski, S. R. 1993, ApJ, 409,  635
\bibitem[Reid et al. (1996)]{rei96}
Reid, I. N., Yan, L., Majewski, S., Thompson, I., Smail, I.
1996, AJ, 112, 1472
\bibitem[Reid et al. (1997)]{rei97}
Reid, I. N., Gizis, J. E., Cohen, J. G., Pahre, M. A.,
Hogg, D. W., Cowie, L., Hu, E., and Songaila, A. 1997, 
PASP, 109, 559
\bibitem[Reid and Gizis (1997)]{rg97} Reid, I. N. and Gizis, J. E.,
AJ, 113, 2246
\bibitem[Saumon et al. (1994)]{sau94} Saumon, D., Bergeron, P., 
Lunine, J. I., Hubbard, W. B., and Burrows, A. 1994, ApJ, 424, 333
\bibitem[Schmidt et al. (1995)]{sch95}
Schmidt, M., Schneider, D. P., and Gunn, J. E.  1995, AJ, 110, 68
\bibitem[Strauss et al. (1999)]{str99}
Strauss, M. A. et al. 1999, ApJ, 522, L61
\bibitem[Tsvetanov et al. (2000)]{tsv00}
Tsvetanov, Z. I. et al.  2000, ApJ, 531, L61
\end{thebibliography}
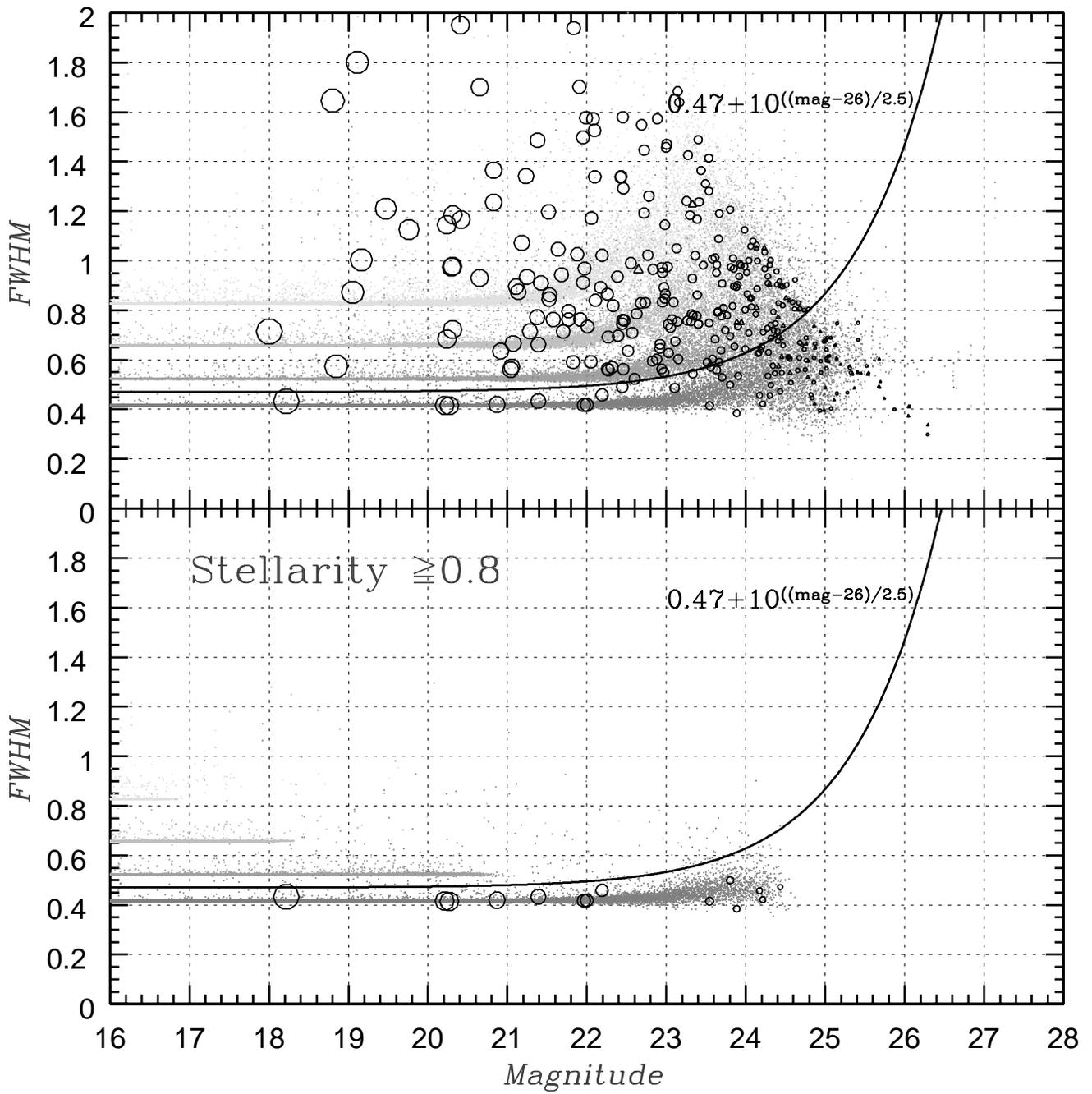
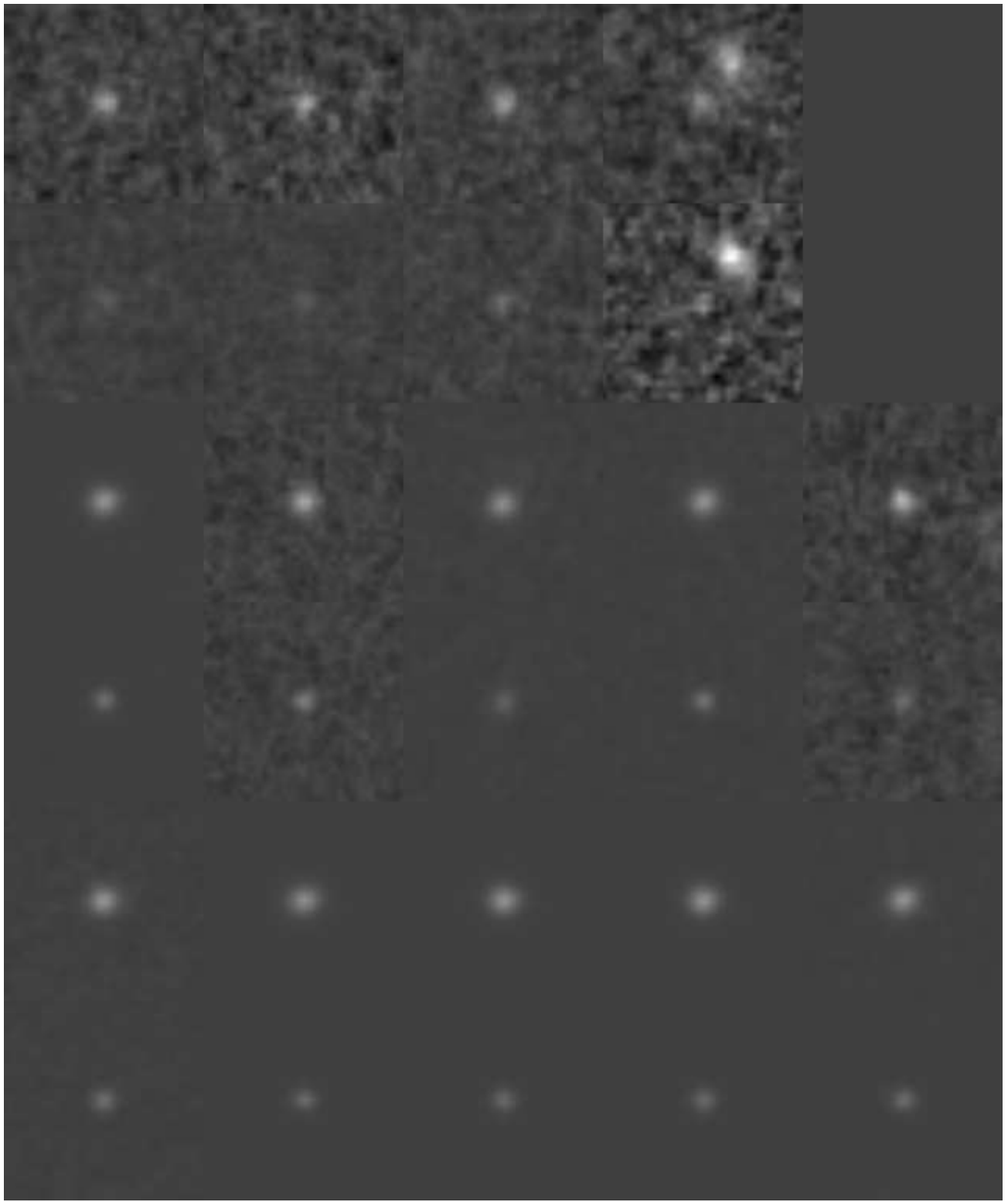
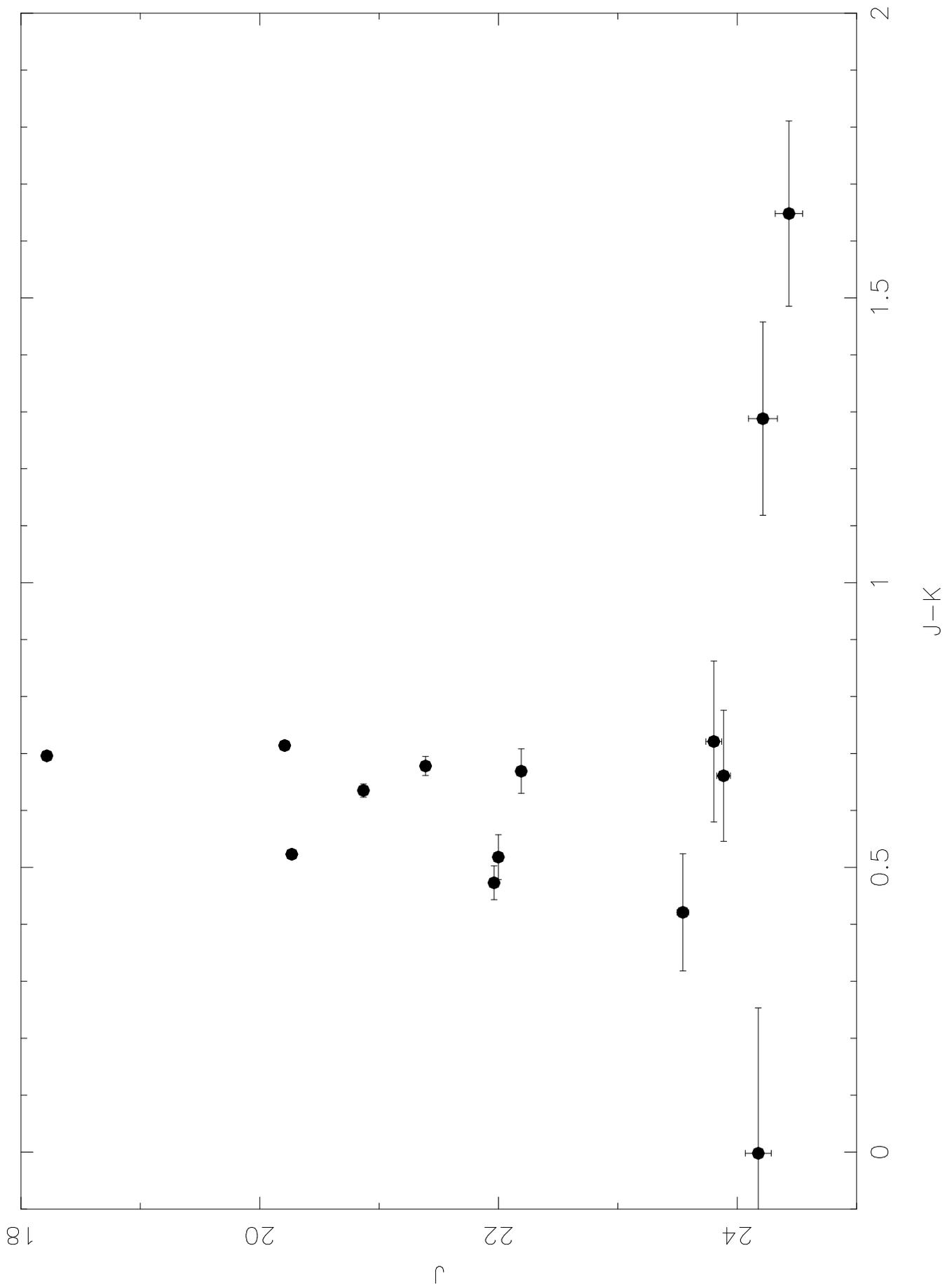
\end{document}